# Assessment of El Niño effects in the North Eurasian regions: predictability of warm and cold winters

I.I. Mokhov

A.M. Obukhov Institute of Atmospheric Physics RAS
Lomonosov Moscow State University
mokhov@ifaran.ru

**Abstract**

The predictability of extreme winter regimes in the regions of Northern Eurasia depending on El Niño phenomena is assessed using data from long-term meteorological observations. The frequency of extremely warm and extremely cold winters is compared for different phases and different types of El Niño events.

**Introduction**

Under global warming with the most rapid climate changes in high latitudes [1,2], a significant increase in weather and climate anomalies in recent decades is noted, in particular in the Russian regions (http://www.meteorf.ru). The number of dangerous meteorological phenomena in Russia has increased by about two dozen events per year since the end of the 20th century [3,4]. Along with summer heat waves, there are winter cold waves. Their manifestation is facilitated by increased tortuosity of atmospheric jet streams under warming with an increase in the probability of intrusions into the middle latitudes of cold northern air or warm air from southern latitudes and the formation of prolonged atmospheric blockings with corresponding intraseasonal temperature anomalies [5].

Under global warming, zonal circulation in the troposphere in mid-latitudes may increase due to the cooling of the stratosphere and mesosphere, which contributes to the strengthening of jet streams. In recent decades, particularly in the Northern Hemisphere (NH), there have been changes in the intensity of the subtropical jet stream, including some strengthening in summer and a stronger weakening in winter. In summer the interannual variability of the jet stream intensity is significantly less than in winter. Features of the connection with changes in the atmospheric jet stream of seasonal blocking activity are noted in [6]. According to estimates [7], about 80% of summer heat waves and about 60% of winter cold waves in the NH are associated with atmospheric blockings [5].

One of the climate problems of recent years is associated with studies of the processes of formation of cold winter conditions over continental regions in the NH (in particular, in Eurasia) under global warming. Their formation is influenced by various processes that determine regional climatic variability against the background of longer-term changes. In particular, numerous studies are devoted to the analysis of the connection between winter temperature anomalies in the middle latitudes with the strongest warming in the Arctic latitudes and negative anomalies in the extent of sea ice in the Arctic basin [1-10].

The occurrence of cold winters in mid-latitude regions in recent decades under warming is associated with previously obtained empirical and model estimates of changes in the conditions for the formation of atmospheric blockings [11-13]. According to estimates obtained in [12] based on long-term data, with near-surface warming of the NH, the characteristic lifetime of atmospheric blockings in mid-latitudes increases, the consequence of which is, in particular, summer droughts and extreme frosts. The noted empirical trend in [11,12] was given a qualitative explanation using a simple model approach (see also [14]). In particular, under warming with a decrease in the interlatitudinal temperature gradient and a weakening of zonal circulation (geostrophic wind speed) in the troposphere of mid-latitudes, an increase in the duration of blockings should be expected [11,12]. In [13], based on numerical simulations using a general circulation climate model, it was found that under warming, due to an increase in the atmospheric $CO_2$ content, the number and total duration of atmospheric blockings in the NH increase with strongest changes in the winter and spring months over the continents and for the European continent. Atlantic sector. This corresponds to an increase in the risk of winter frosts over continental regions, as a manifestation of climate variability under global warming [5].

The formation of weather and climate anomalies and their frequency in mid-latitudes depend on the climatic conditions of the Arctic, in particular on the Arctic atmospheric center of action. Analysis of long-term data and model simulation results indicate a general trend of weakening of the Arctic anticyclone in the process of global warming [15].

In [16], an analysis of the stability of stationary modes necessary for the manifestation of atmospheric blockings was carried out as part of a study of the dynamics of singular vortices on a rotating sphere (see also [17, 18]. In particular, analytical conditions for the stability of such stationary modes were obtained depending on the intensity of the polar vortex. In this regard, one should expect a dependence on the intensity of the Arctic anticyclonic center of action for the duration of atmospheric blockings in the NH.

Against the background of regional features of blocking activity due to long-term climate changes, regional features associated with key modes of natural interannual and interdecadal climate variability are also noted [19–23]. The strongest influence on interannual variability in global surface temperature is associated with the El Niño/Southern Oscillation. The influence of modes of natural climate variability is also manifested in changes of atmospheric centers of action, including Arctic and Siberian anticyclons [15,24] (see also [5]).

This paper presents estimates of the predictability of winter temperature anomalies in the regions of Northern Eurasia depending on the phases and types of El Niño events using long-term meteorological data for recent decades.

**Data analyzed**

Here, the frequency of warm and cold winters, including extremely warm and extremely cold winters, in the North Eurasian regions is estimated for different phases of El Niño [21]. This paper presents estimates for mid-latitude and southern regions in the European and Asian parts of Russia. The analysis involved the characteristics of warm and cold winters based on monthly averaged data for anomalies of the surface air temperature $\delta T$ in January and February for different Russian regions from meteorological observations for the period of 1936–2014 [25] (see also [26]). The ratio of temperature anomalies $\delta T$ in January and February to the standard deviation $\sigma T$ for the period of 1961–1990, i.e., index $I = \delta T/\sigma T$, was analyzed for different regions. Winters range from extremely warm (EW) and extremely cold (EC) to considerably warm (CW) and considerably cold (CC), as well as moderately warm (MW) and moderately cold (MC) winters. To the south of 60°N, EC-winters were characterized by indices $I$ less than –0.9 for the European part of Russia (ER), as well as for the Amur River region and Primorye, and less than –1 for Cisbaikalia and Transbaikalia. At values of the index $I$ between –0.5 and –0.9, winters were characterized as CC; at values of $I$ between –0.5 and 0, as MC. Correspondingly, to the south of 60°N, EW-winters were characterized by indices $I$ larger than 1.0 for the European part of Russia (ER), as well as for the Amur River region and Primorye, and for Cisbaikalia and Transbaikalia. At values of the index *I* between 0.5 and 1.0, winters were characterized as CW; at values of *I* between 0.5 and 0, as MW.

The effects of El-Nino / La-Nina were estimated using their different indices characterized by the sea surface temperature (SST) in the Nino3 (150°–90°W) and Nino4 (160°E–150°W) regions in subequatorial latitudes of the Pacific (ftp://www.coaps.fsu.edu/pub/). The index Nino3 characterizes the El-Nino of the EP-type with significant positive SST anomalies in the eastern equatorial part of the Pacific (Eastern Pacific, EP). The index Nino4 characterizes the El-Nino of the CP-type with significant positive SST anomalies in the central equatorial part of the Pacific (Central Pacific, CP). Phases of El-Nino (*E*) and La-Nina (*L*) were distinguished using five-month moving averaging of values of the SST anomaly. The El-Nino phase (warm phase) and La-Nina phase (cold phase) were determined by the values of SST anomalies not less than 0.5°C or not greater than –0.5°C over six months in succession, respectively. Other cases were characterized as the neutral phase (*N*).

Based on meteorological data indices of winter anomalies according to [25] for the selected 10 regions were used (see also [21,26]): northern regions (to the north from 60ºN - European part of Russia, Western Siberia, Central Siberia, Eastern Siberia; southern regions (to the south from 60ºN) - European part of Russia (SER), Western Siberia, Central Siberia, Lake Baikal basin and Transbaikalia (LBT), Eastern Siberia, Amur River basin and Primorye (ARP).

**Results**

Table 1 (a, b, c) presents the number (*n*) and frequency estimates (*n/n$_\Sigma$*) of extremely warm (EW), extremely and considerably warm (EW + CW) and extremely, considerably and moderately warm (EW + CW + MW), as well as extremely cold (EC), extremely and considerably cold (EC + CC) and extremely, considerably and moderately cold (EC + CC + MC) winters for the SER (a), LBT (b) and ARP (c) to the south from 60ºN at different El Niño phases, characterized by the Nino3 and Nino4 indices (see also [21,26]).

Table 1. The number (*n*) and frequency estimates (*n/n$_\Sigma$*) of extremely warm (EW), extremely and considerably warm (EW + CW) and extremely, considerably and moderately warm (EW + CW + MW), as well as extremely cold (EC), extremely and considerably cold (EC + CC) and extremely, considerably and moderately cold (EC + CC + MC) winters for the SER (a), LBT (b) and ARP (c) to the south from 60ºN at different El Niño phases, characterized by the Nino3 and Nino4 indices.

| (a) | | Southern European Region | | | | | |
|---|---|---|---|---|---|---|---|
| *n/n$_\Sigma$* (1936-2014) | | Warm Winters | | | Cold Winters | | |
| | | EW | EW+CW | EW+CW+MW | EC | EC+CC | EC+CC+MC |
| Nino3 | *N* | 4/44 | 14/44 | 25/44 | 5/44 | 11/44 | 19/44 |
| | *L* | 3/19 | 7/19 | 9/19 | 3/19 | 7/19 | 10/19 |
| | *E* | 1/16 | 6/16 | 11/16 | 0/16 | 1/16 | 5/16 |
| Nino4 | *N* | 3/40 | 13/40 | 21/40 | 3/40 | 9/40 | 19/40 |
| | *L* | 3/18 | 7/18 | 9/18 | 3/18 | 7/18 | 9/18 |
| | *E* | 2/21 | 6/21 | 15/21 | 2/21 | 3/21 | 6/21 |

| (b) $n/n_\Sigma$ (1936-2014) | | Lake Baikal basin and Transbaikalia | | | | | |
|---|---|---|---|---|---|---|---|
| | | Warm Winters | | | Cold Winters | | |
| | | EW | EW+CW | EW+CW+MW | EC | EC+CC | EC+CC+MC |
| Nino3 | N | 4/44 | 17/44 | 22/44 | 5/44 | 11/44 | 22/44 |
| | L | 4/19 | 8/19 | 11/19 | 0/19 | 0/19 | 8/19 |
| | E | 0/16 | 1/16 | 4/16 | 3/16 | 8/16 | 12/16 |
| Nino4 | N | 3/40 | 13/40 | 16/40 | 7/40 | 13/40 | 24/40 |
| | L | 3/18 | 8/18 | 12/18 | 0/18 | 0/18 | 6/18 |
| | E | 2/21 | 3/21 | 9/21 | 1/21 | 6/21 | 12/21 |

| (c) $n/n_\Sigma$ (1936-2014) | | Amur River basin and Primorye | | | | | |
|---|---|---|---|---|---|---|---|
| | | Warm Winters | | | Cold Winters | | |
| | | EW | EW+CW | EW+CW+MW | EC | EC+CC | EC+CC+MC |
| Nino3 | N | 5/44 | 13/44 | 22/44 | 4/44 | 11/44 | 22/44 |
| | L | 2/19 | 10/19 | 11/19 | 1/19 | 3/19 | 8/19 |
| | E | 1/16 | 3/16 | 7/16 | 3/16 | 5/16 | 9/16 |
| Nino4 | N | 3/40 | 10/40 | 17/40 | 2/40 | 10/40 | 23/40 |
| | L | 2/18 | 9/18 | 11/18 | 2/18 | 3/18 | 7/18 |
| | E | 3/21 | 7/21 | 12/21 | 4/21 | 6/21 | 9/21 |

According to Table 1a in SER at *L*-phase the frequency of the EW and (EW + CW) conditions, as well as the EC, (EC + CC) and (EC + CC + MC) conditions, is greater than for the N-phase and E-phase for both types El Niño, characterized by the Nino3 and Nino4 indices. At the same time, the frequency of the (EW + CW + MW) conditions in the *E*-phase is greater than in the *N*-phase and *L*-phase. Moreover, in the E-phase, the probability estimates for the (EW + CW + MW) conditions are more than twice as high as the probability estimates for the (EC + CC + MC) conditions. Of particular note is that SER was never observed during past decades in EC conditions at *E*-phase using the Nino3 index.

In LBT, according to Table. 1b, during the *L*-phase, the frequency of the EW, (EW + CW) and (EW + CW + MW) conditions is greater than during the *N*-phase and *E*-phase for both types of El Niño, characterized by the Nino3 and Nino4 indices. At the same time, the frequency of the EC, (EC + CC) and (EC + CC + MC) conditions is greater in the *E*-phase than in the *N*-phase and *L*-phase when using the Nino3 index, whereas when using the Nino4 index, the frequency of the EC, (EC + CC) and (EC + CC + MC) conditions is greater in the *N*-phase than in the *E*-phase and *L*-phase. It is significant that for LBT in the *L*-phase the EC and (EC + CC) conditions were never observed during past decades when using the Nino3 index. The probability estimate for the (EC + CC + MC) conditions in the *E*-phase for LBT is three times greater than the probability estimate for the (EW + CW + MW) conditions using the Nino3 index. Moreover, the estimate of the probability of the (EC + CC) conditions is 8 times greater than the estimate of the probability of the (EW + CW) conditions, and the EW conditions were never observed during past decades in the *E*-phase. When using the Nino4 index in *N*-phase for LBT, the probability estimate for the (EC + CC + MC) conditions one and a half times greater than the probability estimate for the (EW + CW + MW) conditions, and more than twice as large for the EC conditions than for EW conditions. It should be especially noted that for LBT in the *L*-phase the EC and (EC + CC) conditions were never observed during past decades when using the Nino4 index, as well as when using the Nino3 index. And the probability of (EW + CW + MW) conditions in the *L*-phase is estimated to be twice as high as the probability of (EC + CC + MC) conditions. In the *E*-phase, the probability estimates for the (EC + CC + MC) and (EC + CC) conditions exceed the corresponding probability estimates for the (EW + CW + MW) and (EW + CW) conditions. Moreover, only once in the LBT during 21 El Niño events was an EC conditions and two EW conditions observed.

In ARP, according to Table. 1c, during the *L*-phase, the frequency of the (EW + CW) and (EW + CW + MW) conditions is greater than during the *N*-phase and *E*-phase for both types of El Niño, characterized by the Nino3 and Nino4 indices. Moreover, the frequency of the EC and (EC + CC) conditions is greater in the E-phase than in the *N*-phase and *L*-phase when using both Nino3 and Nino4 indices. At the same time, the frequency of the (EC + CC + MC) conditions is greatest in the E-phase when using the Nino3 index and in the *N*-phase when using the Nino4 index. The frequency of the EW conditions in the *E*-phase was rated highest when using the Nino4 index and lowest when using the Nino3 index. In the *L*-phase, when using the Nino3 index, the probability estimates for the EW, (EW + CW) and (EW + CW + MW) conditions are greater than the corresponding probability estimates for the EC, (EC + CC) and (EC + CC + MC) conditions, and in the *E* -phase – vice versa.

When using the Nino4 index, the probability estimates for the (EW + CW) and (EW + CW + MW) conditions in the ARP in both the *L*-phase and *E*-phase are greater than the corresponding probability estimates for the (EC + CC) and (EC + CC + MC) conditions. At the same time, the estimates of the probability of the EW and EC conditions in the *L*-phase were obtained the same, and in the E-phase the probability of the EC conditions was estimated to be greater.

Quite different temperature anomalies have appeared in recent years in the North Eurasian regions. Figure 1 shows strong regional anomalies of the surface air temperature in Januaries and Februaries in 2015, 2016, 2018, 2019, 2021, 2022, 2023 by GISS data (https://data.giss.nasa.gov/gistemp/). In particular, as noted in [26], over a large territory of Russia in the winter of 2020/2021, including in January and February 2021, significant negative anomalies of surface temperature were detected, while positive temperature anomalies were noted for the Lake Baikal basin and Transbaikalia. According to meteorological observations (climatechange.igce.ru), positive temperature anomalies (relative to 1961-1990) were observed throughout the winter months in the Lake Baikal basin and Transbaikalia. In other large Russian regions (including in the European and Asian parts of Russia, in Western, Central and Eastern Siberia, the Amur River basin and Primorye) and in Russia as a whole, temperature anomalies were negative in at least one of the winter months, including January and February 2021. After 2014, another year, along with 2021, began in the *L*-phase - 2018. During the winter months of 2017/2018, according to meteorological observations, positive temperature anomalies were also observed in the Lake Baikal basin and Transbaikalia during all winter months [26].

Observational data for recent winters confirm the significance of estimates obtained from previously obtained data, in particular for winters in the Lake Baikal basin and Transbaikalia in the *L*-phase. In connection with the *L*-phase developing by the end of 2021, it was noted in [26] that in January-February 2022, in particular, the absence of extreme cold in the Lake Baikal basin and Transbaikalia should be expected. The Roshydromet data (http://www.meteorf.ru) confirmed this (see also Fig. 1 (k, l)). Also, the Roshydromet data (http://www.meteorf.ru) confirm the absence of extreme cold in the Lake Baikal basin and Transbaikalia in the *L*-phase in January-February 2019 and 2023 (see also Fig. 1 (g, h) and Fig. 1 (m, n)).

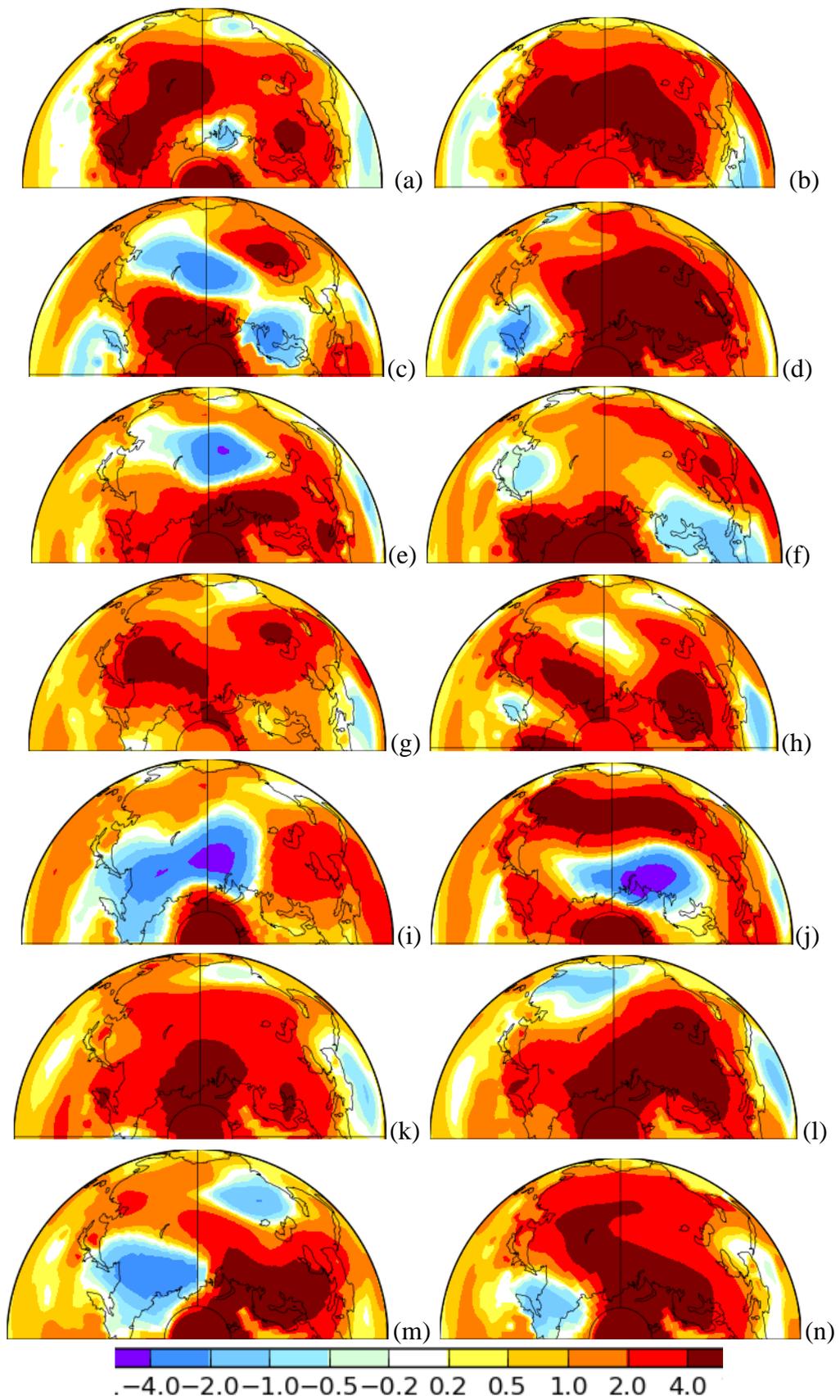

Figure 1. Regional SAT anomalies [K] (relative 1961-1990) in Januaries (a,c,e,g,k,m) and Februaries (b,d,f,,h,j,l,n) from GISS data: 2015 (a,b), 2016 (c,d), 2018 (e,f), 2019 (g,h), 2021 (i,j), 2022 (k,l), 2023 (m,n).

Due to the E-phase forming by the end of 2023, according to the estimates obtained, the probability of a warm winter in SER in January-February 2024 is significantly greater than a cold one, and vice versa in LBT. In this case, we should expect, in particular, the absence of extreme cold conditions in SER and extreme warm conditions in LBT in January-February 2024. The EC conditions in SER and the EW conditions in LBT in the *E*-phase using the Nino3 index have not been observed since 1936, and the (EC + CC) conditions and the (EW + CW) conditions have only been observed once. In ARP, the probability of a cold winter in January-February 2024 is greater than a warm one, particularly when using the Nino3 index. After 2014, the *E*-phase was observed, in particular, in 2016 when using both Nino3 and Nino4 indices, and in 2015 only when using the Nino4 index. According to Fig. 1 (c) negative temperature anomalies were noted in the LBT and ARP area in January 2016. According to Roshydromet data (http://www.meteorf.ru), negative temperature anomalies were noted in the ARP area in February 2016.

**Conclusions**

As noted in [26], obtaining more reliable estimates of regional weather and climate predictability does require more detailed and comprehensive data analysis and ensemble model simulations. However, even the relatively small statistical base of available data and various types of El Niño phenomena makes it possible to identify effects that are significant for obtaining prognostic estimates of regional weather and climate anomalies.

It should be noted that key modes of interannual and interdecadal variability, including El Niño events, and their influence on different regions are changing under global climate change. In particular, the trend of intensification and increase in the frequency of El Niño phenomena under global warming, as noted in [27,28], is associated with an increased risk of their stronger impact on different regions, including the regions of Northern Eurasia [2]. It should also be noted that there is a need to assess the cumulative impact of various modes of climate variability on the predictability of regional weather and climate anomalies [5].

**Acknowledgements**

This study was supported by the Russian Science Foundation (project 23-47-00104) using the results obtained under Agreement No. 075-15-2021-577 of the Ministry of Science and Higher Education of the Russian Federation with A.M. Obukhov Institute of Atmospheric Physics of the Russian Academy of Sciences.